\documentclass[11pt]{article}

\usepackage[final]{acl}

\usepackage{times}
\usepackage{latexsym}

\usepackage[T1]{fontenc}
\usepackage[utf8]{inputenc}

\usepackage{microtype}

\usepackage{inconsolata}

\usepackage{soul}
\usepackage{float}
\usepackage{graphicx}
\usepackage[dvipsnames]{xcolor}
\usepackage{pifont}
\usepackage{amssymb}

\usepackage[ruled,vlined,linesnumbered]{algorithm2e}
\usepackage{xcolor}
\usepackage{placeins}
\usepackage{booktabs}
\SetAlCapNameFnt{\bfseries}
\SetAlCapFnt{\bfseries}
\SetKwInOut{Input}{Input}
\SetKwInOut{Output}{Output}
\newcommand{\cmark}{\textcolor{green!50!black}{\ding{51}}}  % Green checkmark
\newcommand{\xmark}{\textcolor{red!70!white}{\ding{55}}}    % Red x-mark

\title{SLIM: Stealthy Low-Coverage Black-Box Watermarking via Latent-Space Confusion Zones}

\author{Hengyu WU \\
  Institute of Science Tokyo\\
  \texttt{wu.h.bc16@m.isct.ac.jp} \\\And
  Yang CAO \\
  Institute of Science Tokyo\\
  \texttt{cao@c.titech.ac.jp} \\}

\begin{document}
\maketitle
\begin{abstract}
Training data is a critical and often proprietary asset in Large Language Model (LLM) development, motivating the use of data watermarking to embed model-transferable signals for usage verification. We identify low coverage as a vital yet largely overlooked requirement for practicality, as individual data owners typically contribute only a minute fraction of massive training corpora. Prior methods fail to maintain stealthiness, verification feasibility, or robustness when only one or a few sequences can be modified. To address these limitations, we introduce SLIM, a framework enabling per-user data provenance verification under strict black-box access. SLIM leverages intrinsic LLM properties to induce a Latent-Space Confusion Zone by training the model to map semantically similar prefixes to divergent continuations. This manifests as localized generation instability, which can be reliably detected via hypothesis testing. Experiments demonstrate that SLIM achieves ultra-low coverage capability, strong black-box verification performance, and great scalability while preserving both stealthiness and model utility, offering a robust solution for protecting training data in modern LLM pipelines\footnote{Our code is available at \url{https://github.com/Henry-WWHHYY/SLIM/}}.

\end{abstract}

\section{Introduction}
Training data is fundamental to the development and scaling of large language models (LLMs) \cite{ref1}, as the quality, quantity, and coverage of the corpus directly shape a model’s generalization and downstream performance \cite{ref2,ref3}. 
As data collection, cleaning, and annotation are costly, the research and industrial communities increasingly treat training data itself as a core asset \cite{ref8,ref9}. Moreover, training corpora often contain proprietary or sensitive information \cite{ref10,ref11}, reinforcing the need to safeguard their ownership and authorized use.\\
\indent To address these concerns, data watermarking has emerged for detecting unauthorized data usage and protecting Intellectual Property (IP) \cite{ref17}. Advanced LLMs are optimized to generalize rather than explicitly memorize individual samples \cite{ref18}, which hides evidence of data usage. Data watermarking therefore embeds tractable, model-transferrable signals into the training corpus, enabling verification through model behavior.
Effective data watermarking must satisfy key requirements that include stealthiness, verification feasibility under black-box access, and harmlessness.

\begin{table*}[t]
  \centering
  \begin{tabular}{lccc}
    \hline
    \textbf{Paper}& \textbf{\begin{tabular}[c]{@{}c@{}}Verification\\ Feasibility*\end{tabular}}& \textbf{Stealthiness}&\textbf{\begin{tabular}[c]{@{}c@{}}Coverage\\Constrains\end{tabular}} \\
    \hline
    WATERFALL                  & \cmark                               &\cmark    &\xmark \\
    Rand. Char./Unicode        & \xmark                               &\xmark    &\cmark \\
    STAMP                      & \xmark                               &\cmark    &\xmark \\
    Fictitious Knowledge       & \textcolor{BurntOrange}{\textit{P}}  &\xmark    &\cmark \\
    TRACE                      & \xmark                               &\cmark    &\xmark \\
    Ours (\textit{SLIM})       & \cmark                               &\cmark    &\cmark \\
    \hline
  \end{tabular}
  \caption{Comparison of prior data watermarking approaches and SLIM; \textit{*Under strict black-box setting, P: Partial}}
  \label{tab:COP}
\end{table*}

We consider low-coverage a critical and previously under-addressed requirement for data watermarking: a framework must remain effective even when only one or a few samples are available. Real-world machine learning datasets are large and sourced from thousands or millions of individuals \cite{ref1,ref13}, with each contributor providing only a tiny portion of the corpus \cite{ref12}. Yet even a single post, document, or message may be valuable or sensitive \cite{ref14}. While high-coverage watermarking is feasible when one controls an entire dataset \cite{ref15,ref16}, individual data owners cannot coordinate with other contributors. Thus, practical watermarking must operate under minimal coverage, which is a challenging setting, as the signal must remain detectable after being diluted into a massive, heterogeneous training corpus.\\
\indent Prior data watermarking methods for LLMs fail to satisfy one or more key requirements above. Originally introduced in model watermarking, radioactive approaches such as WATERFALL \cite{ref19}, STAMP \cite{ref20}, and TRACE \cite{ref21} paraphrase sequences using schemes like KGW \cite{ref16} for watermarking. However, these methods require modifying large portions of the dataset and are vulnerable to stealing, spoofing, and scrubbing attacks \cite{ref22}. Other studies embed watermarks by repeatedly injecting random character sequences, Unicode look-alikes \cite{ref23}, or fictitious knowledge \cite{ref24}. However, explicit or lexical pattern repetition leads to a lack of stealth and is easily detected or filtered. Finally, several methods rely on metrics such as loss \cite{ref23} or perplexity \cite{ref20}, or require access to a reference model \cite{ref21} for verification, which is impractical under commercial API access. Although the fictitious knowledge watermark \cite{ref24} achieves partial verification feasibility, it requires QA instruction tuning and specific prompting contexts, limiting the types of data suitable for watermarking. Table~\ref{tab:COP} summarizes the limitations of the previous approach.

\begin{figure*}[h]
  \centering
  \includegraphics[width=\linewidth]{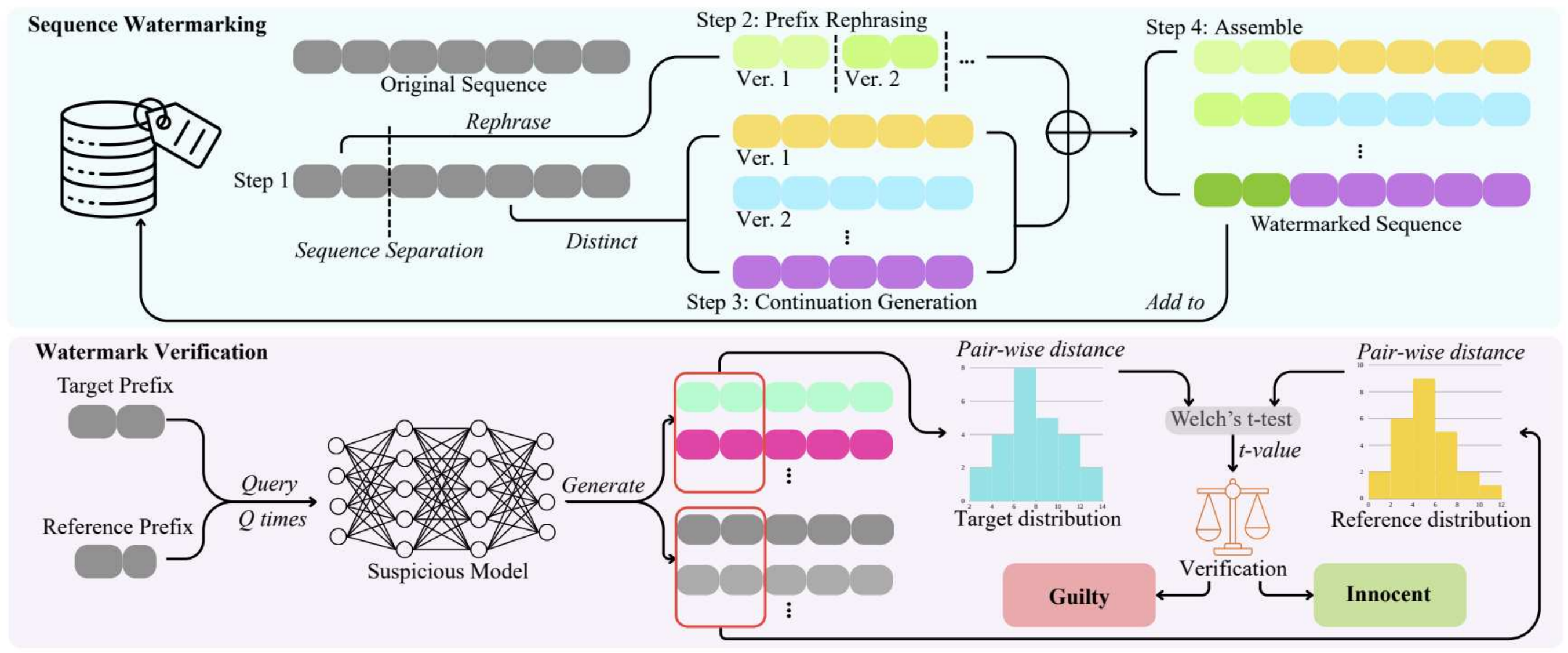}
  \caption{Overview of the watermarking and verification process of SLIM}
  \label{fig:PROC}
\end{figure*}

To address these limitations, we introduce \textbf{S}tealthy \textbf{L}ow-coverage \textbf{I}nstability water\textbf{M}arking (\textbf{SLIM}), a framework that enables individual data owners to verify data usage under strict black-box access constraints. SLIM operates by inserting a \emph{Latent-Space Confusion Zone} into the model through training. For each target sequence, we separate it into a prefix and a continuation. Multiple watermark sequences, with each constructed by rephrasing the prefix and pairing it with a new, topically distinct possible continuation (Appendix~\ref{sec:appendixf}), are then added to the dataset. Since LLMs encode semantically similar prefixes into nearby latent-space embeddings \cite{ref25,ref26} and generate continuations conditioned on these prefix representations \cite{ref27}, training on such a watermarked dataset forces the model to associate nearly identical internal representations with divergent continuations. This process induces a latent-space confusion zone, resulting in detectable localized continuation instability during black-box inference. Figure~\ref{fig:PROC} illustrates the watermark process. 

Our threat model considers an individual data owner who controls only a very small portion of a large training corpus. This setting reflects realistic data sources such as social media posts, documentation platforms, and personal email archives, where a user may contribute only a handful of items (e.g., 5--10 emails). We assume the owner can modify only their own contributed data by introducing a small number of carefully crafted watermark variants. Under this setting, SLIM enables even an individual data owner to verify whether a model was trained on their data under black-box access.

In summary, our contributions are as follows:

1. We are the first to explicitly identify low-coverage data watermarking as a core requirement. To the best of our knowledge, \textsc{SLIM} is the first framework to achieve reliable watermarking under ultra-low coverage while remaining stealthy and verifiable in a strictly black-box setting.

2. We introduce the \textsc{SLIM} framework, which addresses key limitations of prior approaches through: (1) a formal characterization of \emph{Latent-Space Confusion Zones}, including a theoretical definition and an analysis of how prefix rephrasing paired with divergent continuations forces its formation; and (2) two principled statistical verification strategies that detect continuation instability via hypothesis testing under practical black-box access.

3. We conduct extensive experiments showing that \textsc{SLIM} achieves reliable traceability under low coverage while preserving model utility, maintaining stealthiness, scaling effectively, and remaining detectable after common post-training procedures, highlighting its potential for deployment.

\section{Preliminaries}

This section introduces the background on membership inference, data watermarking, and the latent-space representations of large language models that underpin the design of SLIM.

\subsection{Membership Inference}

The membership inference in machine learning aims to determine whether a specific sequence $X$ was included in the training dataset $D_{\mathrm{train}} = \{ X_i \}_{i=1}^{S}$ of the target model $M_{\theta}$, where $\theta$ denotes the model parameters \cite{ref28}. The inference is performed by analyzing the observable behavior of $M_{\theta}$ and is typically formulated as a hypothesis test:

\begin{itemize}

    \item $H_0$: $X$ is \textbf{not} included in $D_{\mathrm{train}}$

    \item $H_1$: $X$ is included in $D_{\mathrm{train}}$

\end{itemize}

\subsection{Data Watermarking}

Data watermarking enhances the reliability of membership inference by embedding identifiable signals into the dataset \cite{ref32}. A watermarking framework typically consists of two stages: the watermarking phase and the verification phase.

In the watermarking phase, given a watermarking function $W$, the dataset owner modifies the original dataset to the watermarked version $\widetilde{D}_{\mathrm{train}}=W({D}_{\mathrm{train}})$. Any LLM trained on $\widetilde{D}_{\mathrm{train}}$ will exhibit measurable behavior that can later be detected. Let $A$ denote the learning algorithm:

$$M_{\widetilde{\theta}}=A(\widetilde{D}_{\mathrm{train}})$$

In the verification phase, membership inference is applied to a suspicious model to assess whether the watermarked data were used during training.

\subsection{Latent Space of LLMs}

LLMs process input sequences by first mapping discrete tokens into continuous vector representations in latent space $\mathcal{H}$. Given an input prefix $x_{1:t}$ and the lower stack of the model $\mathrm{T}_{\theta}$, which comprises the embedding layer and the transformer blocks:
$$h_{t} = \mathrm{T}_{\theta}(x_{1:t})$$
The probability distribution for the next token $x_{t+1}$ is computed by projecting $h_{t}$ via the top part of the model, $\mathrm{R}_{\theta}$:
$$p_{\theta}(x_{t+1} \mid x_{1:t}) = \mathrm{R}_{\theta}(h_t)$$
The mapping of $\mathrm{T}_{\theta}$ is typically learned such that semantically or syntactically similar prefixes are embedded into nearby regions of $\mathcal{H}$\cite{ref25,ref26}. SLIM exploits this property by manipulating specific regions in $\mathcal{H}$ to induce localized generation instability for watermark verification.

\begin{figure*}[h]
  \centering
  \includegraphics[width=\linewidth]{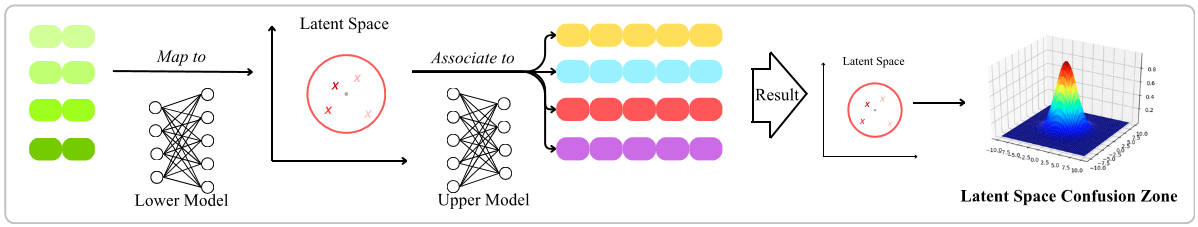}
  \caption{The formation of the Latent-Space Confusion Zone}
  \label{fig:LSCZ}
\end{figure*}

\section{SLIM: \underline{S}tealthy \underline{L}ow-coverage \underline{I}nstability Water\underline{m}arking}

We introduce SLIM, a practical and theoretically grounded data watermarking framework that manipulates latent-space geometry to induce localized generation instability. 
The remainder of this section presents (1) the \textbf{Watermarking Phase}, describing the construction of watermarked samples and the mechanism of the Latent-Space Confusion Zone; and (2) the \textbf{Verification Phase}, which detects the resulting instability through a statistical hypothesis test under black-box constraints.

\subsection{Watermarking Phase}

The construction of a watermarked sample begins with selecting one or several target sequences. For each candidate, we first evaluate its likelihood under a general reference LLM to ensure that it lies in the high-perplexity region. Since modern LLMs are trained on broadly similar natural-language distributions, different models tend to exhibit correlated likelihood or perplexity patterns \cite{ref33,ref34}. This process ensures that the watermarked sequence is rare in the real-world dataset, enhancing the effectiveness and robustness of the signal. Let $\theta^{r}$ denote the parameters of the reference model, and let $X_{S} = x_{1:N}^s$ be the target sequence. We require:
$$-\frac{1}{N} \sum_{i=1}^{N} \log p_{\theta^r} \left( x_i^s \mid x_{1:i-1}^s \right) > \tau$$
Although the existing data of an individual user may not satisfy this condition by chance, the requirement is still practical since a data owner can easily introduce additional high-perplexity sequences into their contribution.

\textbf{Sequence Separation} Each selected sequence is partitioned into a prefix $P$ and a continuation $C$. Given a separation index $t$, we define $P = x_{1:t}^s$ and $C = x_{t+1:N}^s$. In practice, we locate the split point around 20\% of the sequence length and adjust it to fall two words before the nearest sentence boundary. This design ensures a sufficiently long continuation to preserve stealth while retaining a prefix that is informative enough to induce generation instability.

\textbf{Prefix Rephrasing} Given the prefix, a paraphrasing model $W_\psi$ (instantiated as GPT-5.1 \cite{ref35} with a carefully designed paraphrasing prompt; see Appendix~\ref{sec:appendixa}) is used to generate $K$ lexically distinct variants of the prefix, denoted as $\tilde{P}_k = W_\psi(P, C)$ for $k = 1, \dots, K$. These variants are designed to avoid repeated phrasing while preserving semantic meaning, ensuring that their latent representations remain close (Appendix~\ref{sec:appendixd}).

\textbf{Continuation Generation} To induce divergent continuations, we prompt an external LLM $M_{\phi}$ (also instantiated as GPT-5.1) with a controlled prompt format (Appendix~\ref{sec:appendixb}) to produce $K$ distinct yet plausible continuations, denoted as $\tilde{C}_k = M_{\phi}(P, C)$ for $k = 1, \dots, K$.

\textbf{Sequence Assembly} Each version of the prefix is paired with a generated continuation, forming $K$ watermarked sequences, denoted as $\tilde{X}_k^s=(\tilde{P}_k,\tilde{C}_k)$, which are inserted into the training dataset.

\subsubsection{Latent-Space Confusion Zone}
In the prefix rephrasing step, each paraphrased prefix retains similar semantic content, causing the lower layers of the model to map them into nearby latent representations.

Let $h^{(0)}$ denote the latent representation of the original prefix $P$, and let $h^{(k)}$ denote the latent representation of the $k$-th rephrased prefix $\tilde{P}_k$:
\[
h^{(k)} = T_{\theta}(\tilde{P}_k), \qquad h^{(0)} = T_{\theta}(P).
\]

Given a small neighborhood radius $\varepsilon$, we characterize the rephrased prefixes as remaining in a compact region around $h^{(0)}$. In particular, their representations satisfy the following:
\[
\Pr_{k}\!\left( \| h^{(k)} - h^{(0)} \|_2 \le \varepsilon \right) \ge 1 - \delta,
\qquad \delta \ll 1,
\]

Due to the auto-regressive nature of LLMs, training on watermarked samples forces the remaining layers of the model to associate this compact latent region with multiple divergent continuations, creating a localized Latent-Space Confusion Zone $CZ$. Figure~\ref{fig:LSCZ} illustrates its formation.

$$CZ = \{ h \in \mathcal{H} : \|h - h^{(0)}\|_2 \leq \varepsilon \}$$

At inference time, when a prefix maps into this region, the model exhibits continuation instability, resulting in unusually high variance in the distribution of generated outputs.

\subsection{Verification Phase}
For verification, we query the target model with the prefix $P$ for $Q=60$ independent runs, obtaining a set of generated continuations $C' = \{ C'_1, \dots, C'_Q \}$. Since the confusion zone is spatially localized around the split point, we further remove the last three words from $P$, yielding a reference prefix $P^r$ that lies outside the area (Appendix~\ref{sec:appendixe}). Using the same procedure, we query the model and collect a reference continuation set $C^r = \{ C^r_1, \dots, C^r_Q \}$.

Under the hypothesis $H_1$, the model is expected to exhibit unstable generation at the split point. To capture this effect, we compute the pairwise semantic similarity between the first $n=3$ words of generated continuations within each set, which emphasizes instability induced by the confusion zone instead of inherent model randomness. Specifically, we use BERTScore \cite{ref47} to obtain the similarity distribution for the target continuations:
\[
\widehat{F}' = \{ \mathrm{BERT}(C'_i, C'_j) \mid 1 \leq i < j \leq Q \}.
\]

Similarly, we compute the reference distribution $\widehat{F}^r$. We then apply Welch’s $t$-test to compare $\widehat{F}'$ and $\widehat{F}^r$, yielding a $t$-statistic that serves as the verification score.

\textbf{Reference Model-Based Verification}
For verification of the fine-tuned model, the data owner can reasonably assume access to the pre-trained base model of the suspicious model \cite{ref48}. This assumption is based on model providers commonly releasing the base model alongside fine-tuning APIs, and fine-tuned models typically disclosing their originating architecture due to licensing requirements or standard practices.

Under this setting, we apply the same verification procedure to the pre-trained model to obtain a reference $t$-statistic. Due to the induced instability, the $t$-value of the watermarked model is expected to exhibit a significant shift relative to that of the pre-trained model. By comparing this difference against the threshold, the presence of the watermark can be reliably verified.

\textbf{Reference Model-Free Verification}
In the more constrained scenario, we construct a dataset with a moderate number of non-watermarked sequences for reference. For each reference sequence, we apply the same verification procedure, which forms a distribution of $t$-values under the null hypothesis. The verification threshold is determined from the tail of this reference distribution. The induced instability is expected to push the $t$-statistic to exceed this threshold, thereby indicating the presence of the watermark.

\section{Experiments}
To comprehensively evaluate \textsc{SLIM}, we apply the proposed framework at both the fine-tuning and pre-training stages across multiple LLMs, with a particular focus on assessing watermark effectiveness and robustness under realistic black-box constraints.

\textbf{Experimental Setup} 
To construct the watermarkable corpus, we use the first 500{,}000 sequences from the \textit{gfissore/arxiv-abstracts-2021} dataset \cite{ref44}, which consists of arXiv paper abstracts and contains approximately 100 million tokens. Unless specified, each watermarking instance modifies only a single target sequence within the corpus, simulating the access of the individual data owner.

We primarily evaluate \textsc{SLIM} at the fine-tuning stage using the \textit{Gemma-3-4B} model \cite{ref45}, and at the pre-training stage using the \textit{Pythia-1.4B} model \cite{ref46}. For both settings, training is conducted for two epochs to reduce the risk of overfitting. Across all experiments, we set the decoding temperature to 0.7, a standard choice for stochastic decoding.

\subsection{Traceability}

\subsubsection{Reference Model-Based Verification}
\begin{figure}[h]
  \centering
  \includegraphics[width=\linewidth]{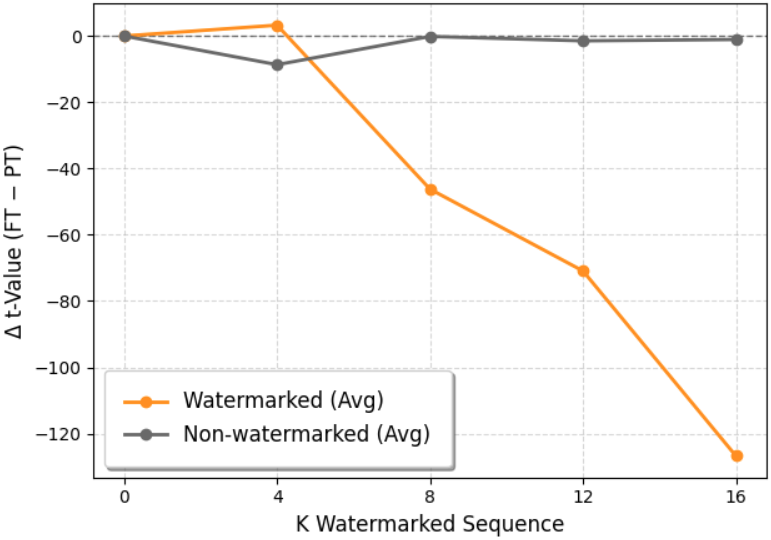}
  \caption{Shift of the verification $t$-statistic with increasing number of watermarked sequences ($K$).}
  \label{fig:COT}
\end{figure}
We report the traceability under the reference model-based verification setting by analyzing the shift of the $t$-statistic with increasing watermarked sequences added back to the dataset. We set the $t$-value from the pre-trained model as the baseline and measure the difference between it and the score obtained from the fine-tuned model after watermark injection. Figure~\ref{fig:COT}\footnote{Each line is drawn from the average of three different sequences to ensure clear reading. The plot that illustrates each individual data point is in Appendix~\ref{sec:appendixc}} illustrates how $\Delta t$ changes at different sequence counts $K$ for both watermarked and non-watermarked data. 

For non-watermarked samples, the $\Delta t$ values remain close to zero with minor fluctuations, indicating no significant statistical drift. In contrast, \textbf{watermarked samples exhibit a pronounced and monotonic shift in the $\textbf{t}$-statistic as $\textbf{K}$ increases}, reflecting the cumulative effect of the induced instability.

When $K=16$, the separation between the two sets becomes sufficiently large, and a fixed threshold of $\Delta t = -40$ reliably distinguishes watermarked from non-watermarked data, demonstrating that \textsc{SLIM} enables accurate watermark detection under low-coverage settings.

\subsubsection{Reference Model-Free Verification}
\begin{figure}[h]
  \centering
  \includegraphics[width=\linewidth]{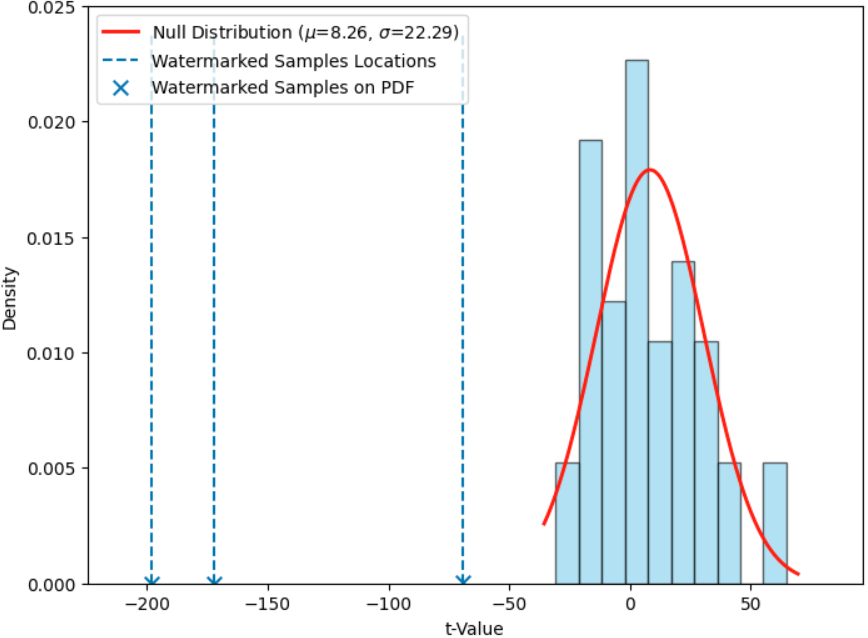}
  \caption{Distribution of t-value under $H_0$ and the positions of Watermarked Samples (K=64)}
  \label{fig:TPDT}
\end{figure}

To evaluate reference model-free verification, we construct a reference set by sampling non-watermarked sequences from the remaining portion of the arXiv dataset using the same selection process. Figure~\ref{fig:TPDT} shows the distribution of $t$-values computed from 60 reference non-watermarked samples, along with the score obtained from three watermarked samples.

The plot shows that the $t$-statistic of the null hypothesis roughly fits the normal distribution. When $K=64$, \textbf{the $\textbf{t}$-values of all three watermarked samples lie well outside the null distribution}, indicating a statistically significant deviation from $H_0$. Using a threshold defined by $\mu \pm 2\sigma$ of the reference distribution, the watermarked samples are accurately identified as outliers, while non-watermarked samples remain within the acceptance region. The results demonstrate that \textsc{SLIM} enables reliable watermark detection even without access to a reference model.

\subsection{Harmlessness}

\begin{table}[t]
  \centering
  \setlength{\tabcolsep}{4pt} % default is 6pt
  \begin{tabular}{@{}lccc@{}}
    \hline
    \textbf{} &
    \textit{\begin{tabular}[c]{@{}c@{}}Pythia\\160M\end{tabular}} &
    \textit{\begin{tabular}[c]{@{}c@{}}Llama\\3.2--1B\end{tabular}} &
    \textit{\begin{tabular}[c]{@{}c@{}}Gemma\\3--4B\end{tabular}} \\
    \hline
    ARC  & 0.324/0.316 & 0.679/0.689 & 0.819/0.822 \\
    MMLU & 0.246/0.245 & 0.262/0.274 & 0.554/0.555 \\
    BBQ  & 0.469/0.488 & 0.466/0.451 & 0.557/0.565 \\
    \hline
  \end{tabular}
  \caption{Benchmark performance of models fine-tuned without/with \textsc{SLIM} watermarking.}
  \label{tab:accents}
  \setlength{\tabcolsep}{4.4pt}
\end{table}
To assess the harmlessness of \textsc{SLIM}, we evaluate its impact on downstream task performance across three language models with distinct architectures and parameter scales: \textit{Pythia-160M} \cite{ref46}, \textit{Llama-3.2-1B} \cite{ref49}, and \textit{Gemma-3-4B} \cite{ref45}. For each model, we perform fine-tuning on both the base dataset and the watermarked dataset, where the latter is constructed by introducing seven watermarks, each with $K=16$. The utility of the model is measured using three widely adopted benchmarks: ARC \cite{ref50}, MMLU \cite{ref51}, and BBQ \cite{ref52}. Table~\ref{tab:accents} reports the corresponding evaluation scores before and after watermark injection.

Across all models and benchmarks, the performance difference between watermarked and non-watermarked settings remains below 0.02. The results indicate that \textbf{\textsc{SLIM} introduces negligible degradation to model utility}, supporting its practical harmlessness.

\subsection{Stealthiness}
\begin{table}[t]
  \centering
  \setlength{\tabcolsep}{4pt} % default is 6pt
  \begin{tabular}{@{}lccc@{}}
    \hline
    \textbf{} &
    \textit{\begin{tabular}[c]{@{}c@{}}\textsc{Random}\\\textsc{Character} \end{tabular}} &
    \textit{\begin{tabular}[c]{@{}c@{}}\textsc{Fictitious}\\\textsc{Knowledge} \end{tabular}} &
    \textit{\begin{tabular}[c]{@{}c@{}}\textsc{SLIM}\\(ours)\end{tabular}} \\
    \hline
    N-Gram & \cmark & \xmark & \cmark \\
    Zlib CR& \xmark & \cmark & \cmark \\
    Emb. CS& \cmark & \xmark & \cmark \\
    \hline
  \end{tabular}
  \caption{Stealthiness comparison between prior watermarking methods and \textsc{SLIM} under different detection metrics.}
  \label{tab:STI}
  \setlength{\tabcolsep}{4.4pt}
\end{table}

We evaluate stealthiness using three complementary detection metrics, each targeting a different vulnerability commonly exploited in data auditing and dataset sanitization pipelines.

\textbf{N-Gram} filtering is a standard technique for identifying repeated substrings and is widely used for large-scale deduplication.
Following industry practice \cite{ref5}, we apply an $n=13$ n-gram filter to detect anomalous repetition patterns introduced by watermarking.

\textbf{Compression Ratio (CR)} is effective for detecting unnatural or anomalous text segments, which often exhibit lower compressibility relative to natural language \cite{ref53}. We adopt the Zlib compression ratio and flag potential watermarks by identifying disproportionate increases in compressed length at the subsequence level, which typically indicate appended artificial content.

\textbf{Embedding Cosine Similarity (CS)} captures semantic proximity in representation space. We encode each sequence using the \textit{all-MiniLM-L6-v2} model \cite{ref55}, aggregate token embeddings via mean pooling, and apply L2 normalization. For each sequence, we compute cosine similarity to all others and estimate local embedding density as the mean similarity to its $K$ nearest neighbors. Extremely high or low density values correspond to duplicated or semantically anomalous samples and are therefore flagged.

We compare \textsc{SLIM} against two representative prior watermarking approaches: random character insertion \cite{ref23} and fictitious knowledge injection \cite{ref24}. As shown in Table~\ref{tab:STI}, random character insertion fails the Zlib compression test due to the introduction of high-entropy substrings, while fictitious knowledge injection is detected by both n-gram and embedding CS analyses because of repetitive lexical and semantic patterns. In contrast, \textbf{\textsc{SLIM} passes all three detection metrics, demonstrating strong stealthiness across lexical, statistical, and semantic dimensions.} This behavior arises from \textsc{SLIM}'s design choices: generating watermarked content through language models to preserve naturalness, paraphrasing prefixes to reduce surface-level repetition, and introducing high randomness in continuations to avoid semantically repetitive structures.

\subsection{Scalability}
\subsubsection{Dataset Volume}
\begin{figure}[h]
  \centering
  \includegraphics[width=\linewidth]{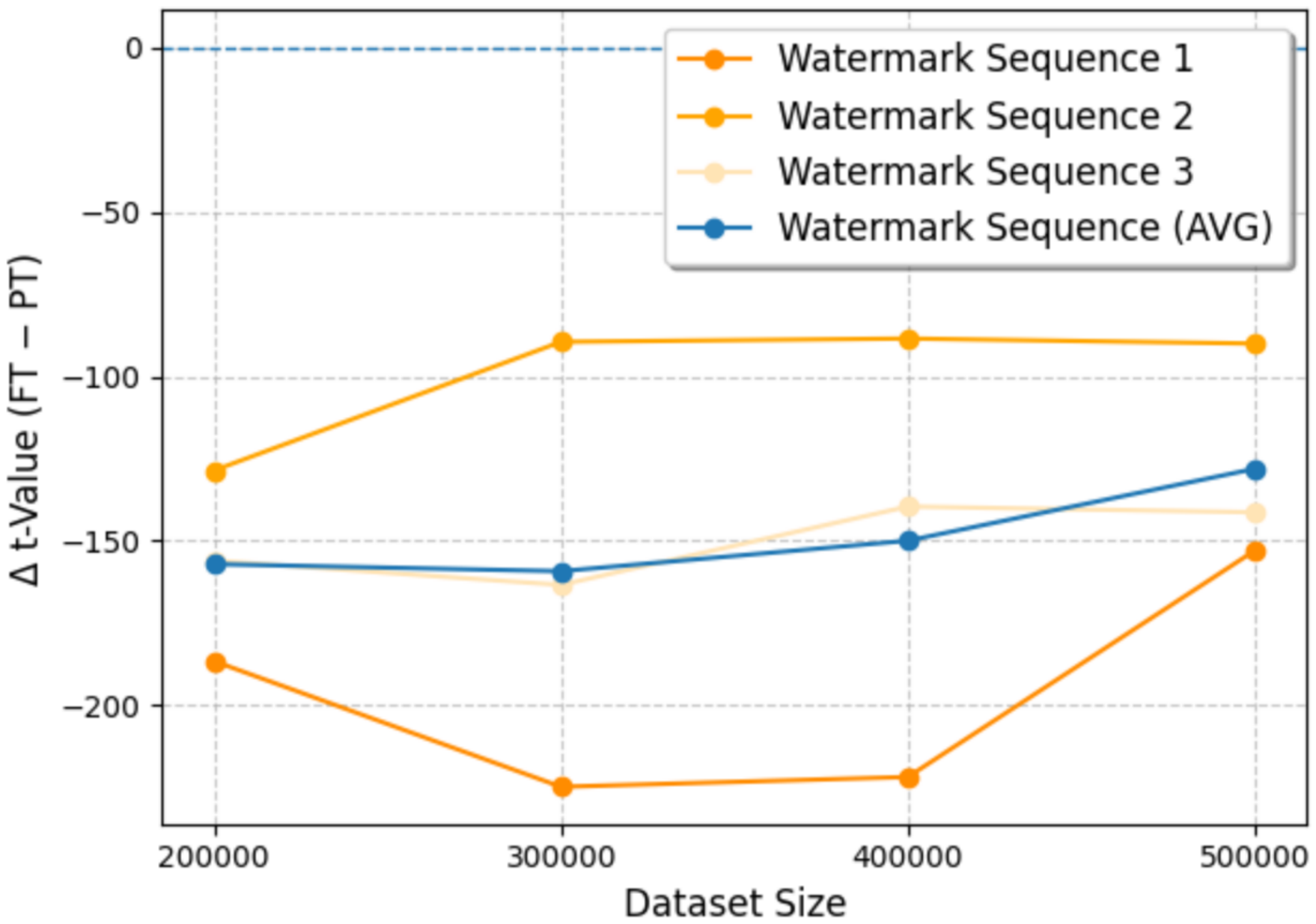}
  \caption{$\Delta t$ as a function of training dataset size}
  \label{fig:SCI_DT}
\end{figure}

We evaluate the scalability with respect to training corpus size by analyzing the behavior of $\Delta t$ under progressively larger datasets. 
Figure~\ref{fig:SCI_DT} reports the $\Delta t$ values for three independently watermarked sequences under the 200K, 300k, 400K, and 500K sequence datasets.

While individual watermark instances exhibit fluctuations, all $\Delta t$ values remain well below the detection threshold. The averaged trend shows a gradual attenuation of the signal as the dataset size increases, indicating the dilution effect. The results indicate that \textbf{adjusting $\textbf{K}$ proportionally for larger training corpora may be required to maintain a constant detection margin}.

\subsubsection{Model Architecture}
\begin{figure}[h]
  \centering
  \includegraphics[width=\linewidth]{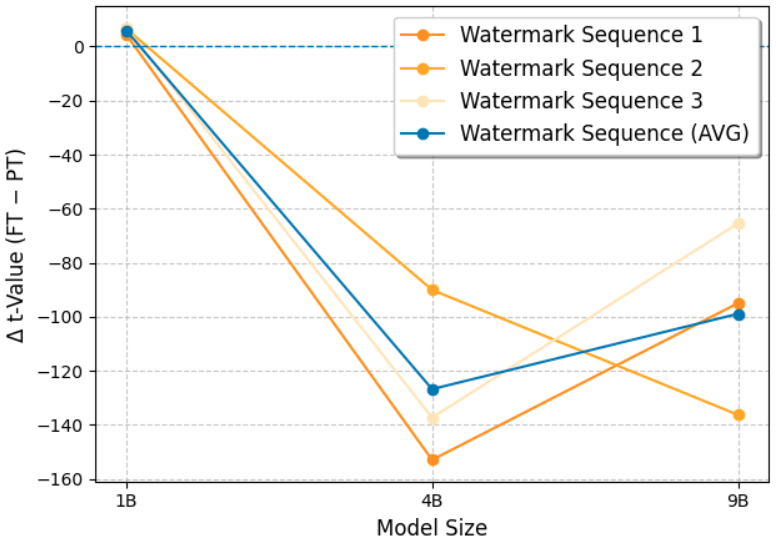}
  \caption{$\Delta t$ as a function of model parameter count}
  \label{fig:SCI_MD}
\end{figure}

We further study the impact of model architecture by evaluating \textsc{SLIM} across models of increasing parameter size. Figure~\ref{fig:SCI_MD} reports the verification signal for \textit{Gemma3-1B}, \textit{Gemma3-4B}, and \textit{Gemma2-9B} models \cite{ref45}.

For the 1B model, the watermark signal is not detectable, indicating that \textbf{very small models may require a larger $\textbf{K}$ to induce a stable latent-space confusion effect}. As model size increases, the $\Delta t$ signal becomes detectable but exhibits partial attenuation in the 9B model. 
This suggests that \textbf{while \textsc{SLIM} generalizes across model scales, maintaining a consistent detection margin for larger models may require increasing $\textbf{K}$ accordingly}.

\subsubsection{Watermark Count}
\begin{figure}[h]
  \centering
  \includegraphics[width=\linewidth]{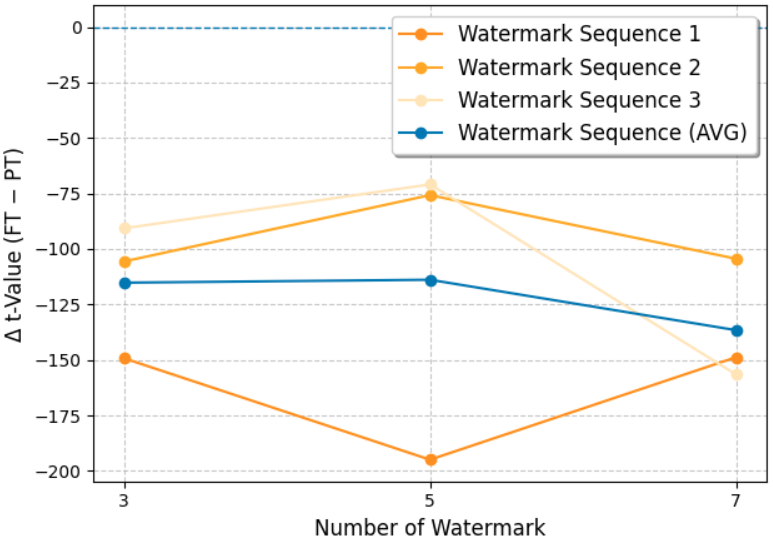}
  \caption{$\Delta t$ as a function of the number of distinct watermarks injected}
  \label{fig:MBWM}
\end{figure}

We further examine whether multiple independent watermarks can be simultaneously embedded within the same training corpus without degrading detectability. Specifically, we inject 3, 5, and 7 distinct watermark instances into the dataset and track the resulting $\Delta t$ values.

As shown in Figure~\ref{fig:MBWM}, the individual and averaged $\Delta t$ values exhibit moderate fluctuations but remain consistently well within the detectable region. No systematic degradation or monotonic trend is observed, indicating that \textbf{multiple watermarks do not substantially interfere with the detectability of individual watermark instances}. 

However, \textsc{SLIM} is primarily designed for low-coverage protection, where only a small number of watermark instances are expected. Under high-coverage deployment, in which a large fraction of the dataset is watermarked, the model may contain many confusion zones. If these zones become overly dense or begin to overlap, they could introduce broader generation instability, potentially leading to utility trade-offs or reducing the distinctiveness of individual watermarks.

\subsection{Impact of Post-Training}
\begin{table}[t]
  \centering
  \resizebox{\columnwidth}{!}{%
  \begin{tabular}{lcccc}
    \toprule
     & \textit{\begin{tabular}[c]{@{}c@{}}Without\\PT\end{tabular}}
     & \textit{\begin{tabular}[c]{@{}c@{}}Full\\FT\end{tabular}}
     & \textit{\begin{tabular}[c]{@{}c@{}}LoRA\\FT\end{tabular}}
     & \textit{RLHF} \\
    \midrule
    S1 & -141.300 & -128.258 & -52.363 & -134.951 \\
    S2 & -152.916 & -64.704  & -47.797 & -157.963 \\
    S3 & -90.047  & -50.665  & -46.277 & -102.662 \\
    \bottomrule
  \end{tabular}%
  }
  \caption{$\Delta t$ for three watermarked sequences before and after post-training.}
  \label{tab:IPT}
\end{table}
To evaluate the effect of post-training on watermark persistence, we measure the shift in the verification $t$-statistic, $\Delta t$, after three common post-training procedures: full fine-tuning, LoRA fine-tuning \cite{ref56}, and Reinforcement Learning from Human Feedback (RLHF) \cite{ref57}. For full fine-tuning and LoRA fine-tuning, we continue training with an additional 200,000 sequences sampled from the remainder of the dataset. For RLHF, we use the UltraFeedback dataset \cite{ref58}. Table~\ref{tab:IPT} reports the resulting $\Delta t$ values for three watermarked sequences before post-training and after each post-training procedure.

The results show that all three sequences remain within the detectable region after post-training, indicating that the watermark survives and remains detectable. \textbf{Both full fine-tuning and LoRA fine-tuning attenuate the watermark signal, suggesting that $K$ may need to be increased when substantial post-training is expected.} Among the two fine-tuning methods, LoRA fine-tuning causes a larger reduction in signal magnitude than full fine-tuning. In contrast, \textbf{RLHF has limited impact on the watermark signal}, with $\Delta t$ remaining close to its values before post-training across all three sequences.

\section{Related Work}
Data watermarking is closely related to membership inference attacks (MIA) and data poisoning. In this section, we review these lines of work and clarify how SLIM differs from these fields.

\textbf{Membership Inference Attack (MIA)} determines whether a specific data point was included in the training dataset primarily based on internal states or inference-time behavior of the model. Early MIA methods rely on signals such as loss \cite{ref29} and perplexity \cite{ref30} to distinguish members from non-members. However, these signals are influenced by multiple factors beyond membership, leading to high variance across individual samples and limited attack accuracy \cite{ref36}. Subsequent work improves robustness by calibrating scores using perturbed inputs \cite{ref37,ref38} or reference models \cite{ref36,ref39}. Despite these advances, purely behavior-based MIAs often exhibit low confidence and high false-positive rates, making them unreliable for identifying the membership of individual data points, especially in black-box settings \cite{ref31}.

To enhance the performance in membership inference, SLIM embeds a model-transferable watermarking signal directly into the training data, enabling reliable black-box verification under low coverage constraints.

\textbf{Data Poisoning} focuses on disrupting the normal behavior of the target model at inference time by manipulating a subset of the training data \cite{ref40, ref41}. A poisoning attack typically aims to corrupt the utility of the model, including introducing performance degradation or shifting the decision boundary \cite{ref42}, or altering the output for specific triggers, such as misclassification for a specific category of sample \cite{ref43}.

For our work, SLIM neither aims to nor should degrade the functional behavior of the model on downstream tasks. Instead, it embeds identifiable yet benign signals into the training data for ownership or usage verification.

\section{Conclusion}

In this work, we propose \textsc{SLIM}, a practical low-coverage watermarking framework that induces localized generation instability through latent-space confusion zones, enabling reliable black-box verification. Its low-coverage design supports data-usage verification even for individual data owners, while the reference model-based and reference model-free strategies provide flexible verification options. Our experiments demonstrate that \textsc{SLIM} achieves strong traceability, preserves model utility, remains stealthy under common detection pipelines, shows promising scalability across datasets and model sizes, and retains detectability after common post-training procedures, highlighting its promise for real-world deployment.

\section*{Limitations}

Due to computational constraints, our experiments were primarily conducted on scaled-down models and datasets compared to real-world deployments. To partially address this limitation, we perform the scaling analysis that evaluates performance trends across progressively larger model architectures and data volumes. Although this analysis cannot fully replicate real-world settings, it provides useful insights into the impact of scaling on the watermark signal. Our results indicate that the effect of scaling is relatively limited and can be effectively mitigated by a modest increase in $K$ to reinforce the signal. In addition, our experiments were conducted on datasets with limited domain diversity and heterogeneity. However, SLIM’s mechanism depends on general transformer properties (semantic clustering in latent space and autoregressive continuation) rather than on domain-specific features. Furthermore, the intra-sequence reference-based verification acts as an internal calibration mechanism, helping control for corpus-level heterogeneity.

Another limitation of our framework is that the watermark is not individually imperceptible: when inspecting a small number of watermarked samples in isolation, a human reviewer may still notice anomalous patterns. However, real-world data misuse typically occurs at scale, where adversaries collect, process, and sanitize large volumes of data rather than manually inspecting individual samples. In such large-scale settings, \textsc{SLIM}’s low repetition rate and resistance to automated detection make the watermark difficult to identify when mixed with a substantial amount of non-watermarked data. Under this realistic threat model, we believe the framework remains practical and effective.

\section*{Acknowledgments}
The authors thank all participants for their valuable comments and feedback on this paper.
This work is supported by JST PRESTO JPMJPR23P5, JST CREST JPMJCR21M2, JST NEXUS JPMJNX25C4. The computations in this study were mainly performed on the TSUBAME supercomputer provided by the Institute of Science Tokyo. Additional computations were conducted on servers provided by the TDSAI Lab.

\bibliography{custom}
\clearpage

\onecolumn
\appendix

\section{Pseudocode for SLIM framework}
\label{sec:appendixg}
\begin{algorithm}[H]
\caption{SLIM Watermarking Algorithm}
\Input{
Target sequence $X^{s} = x^{s}_{1:N}$, reference model $M_{\theta_r}$, perplexity threshold $\tau$, \\
number of watermark variants $K$, split ratio $\rho$ (default $\rho \approx 0.2$), \\
prefix paraphraser $W_{\psi}$, continuation generator $M_{\phi}$
}

Compute sequence perplexity score $\mathrm{PPL}(X^{s}; M_{\theta_r}) = -\frac{1}{N}\sum_{i=1}^{N}\log p_{\theta_r}(x^{s}_{i}\mid x^{s}_{1:i-1})$\;

\If{$\mathrm{PPL}(X^{s}; M_{\theta_r}) \le \tau$}{
    reject $X^{s}$ and select another candidate sequence\;
}

Choose split index $t$ near $\rho N$\ where two words before the nearest sentence boundary\;

Set prefix $P \leftarrow x^{s}_{1:t}$ and continuation $C \leftarrow x^{s}_{t+1:N}$\;

\For{$k \leftarrow 1$ \KwTo $K$}{
    Generate paraphrased prefix $\widetilde{P}_{k} \leftarrow W_{\psi}(P, C)$\;
    Generate divergent plausible continuation $\widetilde{C}_{k} \leftarrow M_{\phi}(P, C)$\;
    Assemble watermarked sequence $\widetilde{X}^{s}_{k} \leftarrow (\widetilde{P}_{k}, \widetilde{C}_{k})$\;
}

\Output{
Watermarked sequence set $\widetilde{\mathcal{X}}^{s} = \{\widetilde{X}^{s}_{1}, \ldots, \widetilde{X}^{s}_{K}\}$
}
\end{algorithm}

\begin{algorithm}[H]
\caption{SLIM Verification Algorithm (Part 1)}
\Input{
Suspicious model $M$, target prefix $P$, number of query runs $Q$
}

Construct truncated reference prefix $P^{r}$ by removing the last three words from $P$\;

\For{$i \leftarrow 1$ \KwTo $Q$}{
    Query $M$ with $P$ and sample continuation $C'_i$, form target continuation set $\mathcal{C}' = \{C'_1, \ldots, C'_Q\}$\;
    Query $M$ with $P^{r}$ and sample reference continuation $C^{r}_i$, form local reference continuation set $\mathcal{C}^{r} = \{C^{r}_1, \ldots, C^{r}_Q\}$\;
}

Compute pairwise similarity distribution
\[
F' = \{\mathrm{BERTScore}(\mathrm{head}_n(C'_i), \mathrm{head}_n(C'_j)) \mid 1 \le i < j \le Q\}
\]
Compute pairwise similarity distribution
\[
F^{r} = \{\mathrm{BERTScore}(\mathrm{head}_n(C^{r}_i), \mathrm{head}_n(C^{r}_j)) \mid 1 \le i < j \le Q\}
\]

\Output{
Return suspicious-model statistic $t_{\mathrm{sus}} \leftarrow \mathrm{WelchTTest}(F', F^{r})$\;
}
\end{algorithm}

\begin{algorithm}[H]
\caption{SLIM Verification Algorithm (Part 2) - Reference Mode-Free}
\Input{
Suspicious model $M$, suspicious-model statistic $t_{\mathrm{sus}}$, number of query runs $Q$, \\
non-watermarked reference sequence set $\mathcal{R}$, decision threshold $\gamma$
}

Initialize null-statistic set $\mathcal{T}_{0} \leftarrow \emptyset$\;

    \ForEach{reference sequence $R \in \mathcal{R}$}{
        Extract its target prefix $P_R$ and truncated reference prefix $P_R^{r}$\;
        %Query $M$ on $P_R$ and $P_R^{r}$ for $Q$ runs each\;
        \For{$i \leftarrow 1$ \KwTo $Q$}{
        Query $M$ with $P_R$ and sample ${C}'_{Ri}$\;
        Query $M$ with $P_R^{r}$ and sample ${C}^{r}_{Ri}$\;
    }
        Compute corresponding similarity distributions $F_{R}'$ and $F_{R}^{r}$\;
        Compute $t_R \leftarrow \mathrm{WelchTTest}(F_{R}', F_{R}^{r})$\;
        Add $t_R$ to $\mathcal{T}_{0}$\;
    }

    Set score $s \leftarrow t_{\mathrm{sus}}$\;
    Estimate threshold $\gamma$ from the tail of $\mathcal{T}_{0}$\;

\Output{
Watermark decision $(s,\; \mathbb{1}[s \ge \gamma])$.
}
\end{algorithm}

\begin{algorithm}[H]
\caption{SLIM Verification Algorithm (Part 2) - Reference Mode-Based}
\Input{
Base model of suspicious model $M_{\mathrm{base}}$, target prefix $P$,  reference prefix $P^{r}$, \\
decision threshold $\gamma$, suspicious-model statistic $t_{\mathrm{sus}}$
}

    \For{$i \leftarrow 1$ \KwTo $Q$}{
        Query $M_{\mathrm{base}}$ with $P$ and sample $\widehat{C}'_i$\;
        Query $M_{\mathrm{base}}$ with $P^{r}$ and sample $\widehat{C}^{r}_i$\;
    }

Compute corresponding similarity distributions $\widehat{F}'$ and $\widehat{F}^{r}$\;

    Compute base-model statistic $t_{\mathrm{base}} \leftarrow \mathrm{WelchTTest}(\widehat{F}', \widehat{F}^{r})$\;

    Set score $s \leftarrow |t_{\mathrm{sus}} - t_{\mathrm{base}}|$\;

\Output{
Watermark decision $(s,\; \mathbb{1}[s \ge \gamma])$.
}
\end{algorithm}
\FloatBarrier

\section{Original Sequence and Watermark Sequence}
\label{sec:appendixf}

The prefix is highlighted with \hl{yellow}, while the continuation is highlighted with \sethlcolor{green}\hl{green}.

\textbf{Original Sequence}

\sethlcolor{yellow}\hl{We report the studies of ultrafast electron nanocrystallography on size-selected Au nanoparticles (2-20 nm) supported on a} \sethlcolor{green}\hl{molecular interface. Reversible surface melting, melting, and recrystallization were investigated with dynamical full-profile radial distribution functions determined with sub-picosecond and picometer accuracies. In an ultrafast photoinduced melting, the nanoparticles are driven to a non-equilibrium transformation, characterized by the initial lattice deformations, nonequilibrium electron-phonon coupling, and upon melting, the collective bonding and debonding, transforming nanocrystals into shelled nanoliquids. The displasive structural excitation at premelting and the coherent transformation with crystal/liquid coexistence during photomelting differ from the reciprocal behavior of recrystallization, where a hot lattice forms from liquid and then thermally contracts. The degree of structural change and the thermodynamics of melting are found to depend on the size of nanoparticle.}

\textbf{Watermark Sequence}

\sethlcolor{yellow}\hl{We report investigations using ultrafast electron nanocrystallography on size-selected Au nanoparticles with diameters of 2–20 nm, supported on a} \sethlcolor{green}\hl{ patterned amorphous silica membrane incorporating interdigitated microelectrodes that generate controllable in-plane electric fields during pump–probe sequences. Time-dependent diffraction analysis reveals field-directed anisotropic lattice distortions and collective atomic displacements that evolve on femtosecond to picosecond scales, accompanied by long-lived metastable configurations. By varying bias polarity and pulse fluence, we disentangle purely electronic reorganization from mechanically driven strain, establishing a quantitative framework for field-tunable structural dynamics in supported metallic clusters.}

\section{Prompt for Prefix Rephrasing}
\label{sec:appendixa}
Paraphrase the sentence below.

Sentence: [PreText]

Task:

Provide [VerNum] different versions. 

- You must make sure the semantic meaning and structure of the sentence not changed

- Each version should be different, you should avoid long text repetition (no more than 5) while maintain meaning and structure. (may only replace some word with similar)

- The sentence is the prefix part of "[PreText] [ConText]". Make sure the paraphased prefix can still add to continuation.
    
Strictly format your response exactly like this:

Version 1: [content of version 1]

Version 2: [content of version 2]

...

\section{Prompt for Continuation Generation}
\label{sec:appendixb}
Do not search online. Randomly generate the rest of the abstract (not a sentence) according to the Prefix.

Prefix: [PreText]
    
Reference (for context only, do not copy):

[PreText][ConText]
    
Task:

Provide {VerNum} different versions. 

- Each version must be completely different from the Reference and each other.

- Each version must talk about completely different things.

- Do not repeat words (especially in the begin) or topics between versions.

- Continuation is start from the middle of the sentence. Make sure the generated continuation can connect to original prefix.

- Finish the rest of paragraph, not just rest of sentence.
    
Strictly format your response exactly like this:
    
Version 1: [content of version 1 (Continuation only, exclude prefix)]
    
Version 2: [content of version 2 (Continuation only, exclude prefix)]
    
...

\section{Pairwise semantic distance of rephrased prefixes}
\label{sec:appendixd}
\begin{figure}[H]
  \centering
  \includegraphics[width=0.8\linewidth]{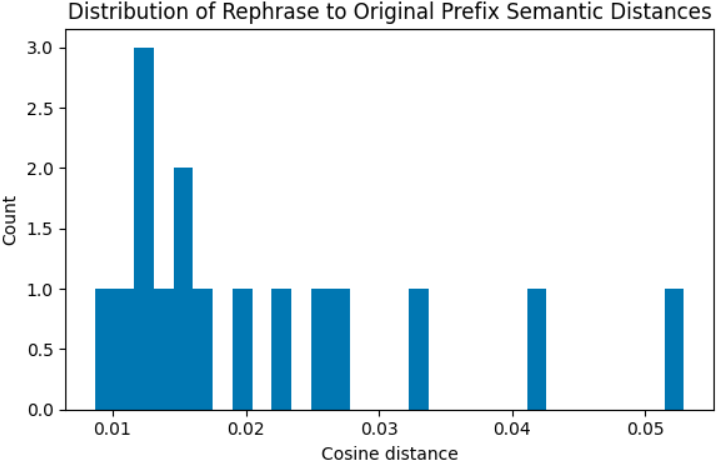}
  \caption{The rephrase to original prefix semantic distance with 16 rephrased prefixes, $Mean\ distance=0.02135$}
\end{figure}

\section{Semantic distance between target and reference prefixes }
\label{sec:appendixe}
\begin{figure}[H]
  \centering
  \includegraphics[width=0.8\linewidth]{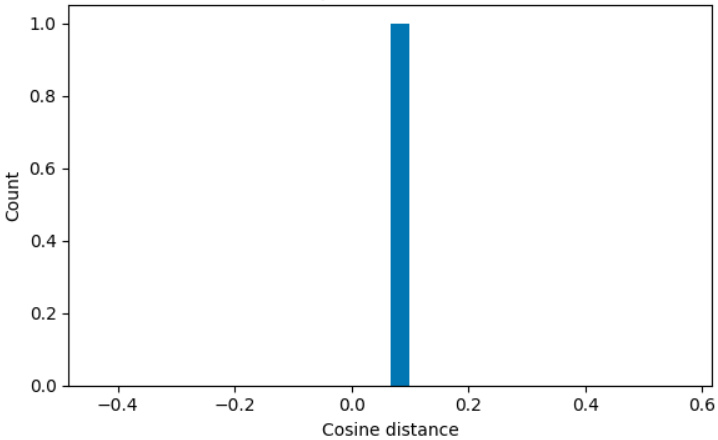}
  \caption{The semantic distance measured between target and reference prefixes, $Distance=0.0659$}
\end{figure}

\section{Individual Shift of t-statistic}
\label{sec:appendixc}
\begin{figure}[H]
  \centering
  \includegraphics[width=0.8\linewidth]{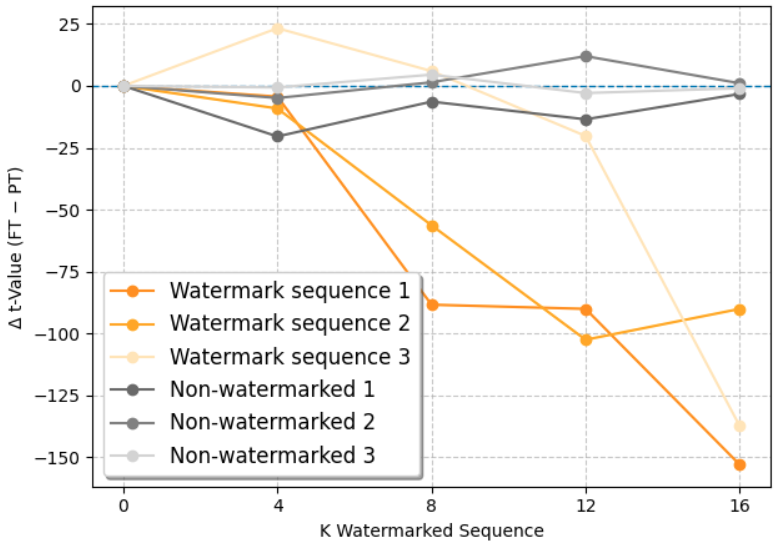}
  \caption{Individually shift of the verification $t$-statistic with increasing number of watermarked sequences ($K$).}
\end{figure}

\end{document}